\documentclass[aps,prx,twocolumn,showpacs,amsmath,amssymb,longbibliography,superscriptaddress]{revtex4-1} 
\usepackage{xcolor}
\usepackage{graphicx}
\usepackage{SIunits}
\usepackage{placeins}

\definecolor{darkblue}{rgb}{0,0,0.6}
\definecolor{darkred}{rgb}{0.6,0,0}
\usepackage[colorlinks=true,urlcolor=darkblue,citecolor=darkblue,linkco
lor=darkred,hyperfootnotes=false]{hyperref}

\newcommand{\Ai}{\mathrm{Ai}}

\begin{document}



\title{Crumples as a generic stress-focusing instability in confined sheets}
\author{Yousra Timounay}
\email{ytimouna@syr.edu}
\affiliation{Department of Physics, Syracuse University, Syracuse, NY 13244}
\affiliation{BioInspired Syracuse: Institute for Material and Living Systems, Syracuse University, Syracuse, NY 13244}
\author{Raj De}
\affiliation{Department of Physics, Syracuse University, Syracuse, NY 13244}
\affiliation{BioInspired Syracuse: Institute for Material and Living Systems, Syracuse University, Syracuse, NY 13244}
\author{Jessica L. Stelzel}
\affiliation{Department of Physics, Syracuse University, Syracuse, NY 13244}
\author{Zachariah S. Schrecengost}
\affiliation{Department of Physics, Syracuse University, Syracuse, NY 13244}
\affiliation{BioInspired Syracuse: Institute for Material and Living Systems, Syracuse University, Syracuse, NY 13244}
\author{Monica M. Ripp}
\affiliation{Department of Physics, Syracuse University, Syracuse, NY 13244}
\affiliation{BioInspired Syracuse: Institute for Material and Living Systems, Syracuse University, Syracuse, NY 13244}
\author{Joseph D. Paulsen}
\email{jdpaulse@syr.edu}
\affiliation{Department of Physics, Syracuse University, Syracuse, NY 13244}
\affiliation{BioInspired Syracuse: Institute for Material and Living Systems, Syracuse University, Syracuse, NY 13244}

\begin{abstract}
Thin elastic solids are easily deformed into a myriad of three-dimensional shapes, which may contain sharp localized structures as in a crumpled candy wrapper, or have smooth and diffuse features like the undulating edge of a flower. 
Anticipating and controlling these morphologies is crucial to a variety of applications involving textiles, synthetic skins, and inflatable structures. 
Here we show that a ``wrinkle-to-crumple'' transition, previously observed in specific settings, is a ubiquitous response for confined sheets. 
This unified picture is borne out of a suite of model experiments on polymer films confined to liquid interfaces with spherical, hyperbolic, and cylindrical geometries, which are complemented by experiments on macroscopic membranes inflated with gas. 
We use measurements across this wide range of geometries, boundary conditions, and lengthscales to quantify several robust morphological features of the crumpled phase, and we build an empirical phase diagram for crumple formation that disentangles the competing effects of curvature and compression. 
Our results suggest that crumples are a generic microstructure that emerge at large curvatures due to a competition of elastic and substrate energies. 
\end{abstract}

\maketitle

\section{Introduction}

When unrolling plastic wrap, handling a large flimsy poster, or watching a fluttering flag, we witness a multitude of deformations available to thin sheets. 
In some cases the shapes are smooth and diffuse like the undulating edge of a flower \cite{Sharon02,Klein07,Gemmer16}, while in others they are sharp and localized, like the ridges and corners in a crumpled piece of paper \cite{Lobkovsky95,Cerda98,Blair05,Cambou11}. 
Although much progress has been made to describe a wide range of deformations and patterns, a general understanding of the transition from smooth to sharp topographies under featureless confinement remains a major challenge. 
Such an understanding promises broad practical implications from controlling surface patterning through buckling \cite{Rodriguez-Hernandez15,Wang16} to anticipating material degradation due to the focusing of stresses at elastic singularities \cite{Tallinen09,Gurmessa13,Gottesman15}. 

Many of these rich and complex morphologies stem from a basic consideration: A sufficiently thin sheet prefers to minimize costly stretching deformations in favor of low-energy bending. 
For sheets constrained to planar or gently curved topographies, wrinkles are an effective method for relaxing compressive stresses while minimizing out-of-plane displacements \cite{Bowden98,Cerda03,Davidovitch11,Bella17}. 
Wrinkles can even allow an initially planar sheet to hug the contour of a doubly-curved geometry, such as a sphere or saddle, with negligible stretching \cite{Hohlfeld15,Vella15,Vella15a,Paulsen16,Davidovitch19,Tobasco19}. 
Yet, under sufficiently strong confinement, a sheet may concentrate strain energy along localized ridges or singular vertices to lower the total elastic energy \cite{Kramer97,Cerda98,Witten07}. 
In this article we study the transition from smooth to sharp deformations in general geometries, and we find a common response whereby wrinkles are replaced at large imposed curvatures by a generic buckling motif, termed ``crumples". 

\begin{centering}
\begin{figure*}[t]
\includegraphics[width=16.2cm]{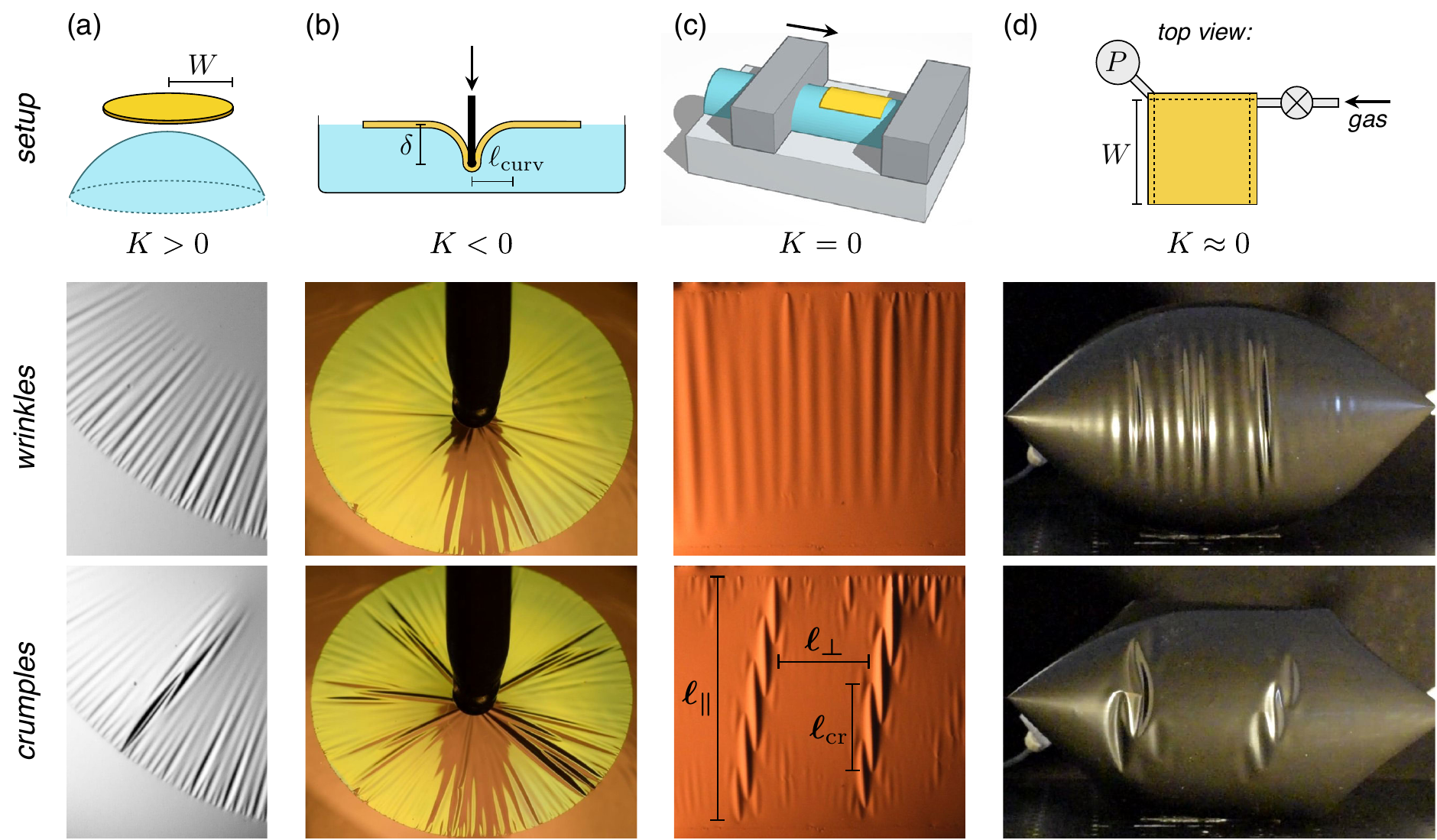} 
\caption{
\textbf{Wrinkle-to-crumple transition in a wide range of geometries, boundary conditions, and system sizes.} 
The overall Gaussian curvature, $K$, imposed on the relevant portion of the sheet is given below each schematic. 
\textbf{(a)} A circular polystyrene sheet ($E=3.4$ GPa) of radius $W=1.5$ mm and thickness $t=77$ nm on a liquid meniscus forms crumples at large droplet curvature \cite{King12}. 
Images adapted from Ref.~\citealp{King13} with permission. 
\textbf{(b)} A polystyrene sheet floating on a liquid bath forms crumples when indented beyond a threshold depth. 
Here, $W=11$ mm and $t=436$ nm. 
\textbf{(c)} Central portion of a rectangular polystyrene film of thickness $t=157$ nm, width $W=3.3$ mm, and length $9.7$ mm on a water meniscus that is uniaxially compressed between two barriers. 
Wrinkles form when the meniscus is flat or gently curved; crumples form when the imposed curvature along the wrinkle crests is sufficiently large. 
\textbf{(d)} A square polyethylene bag ($t=102$ $\mu$m, $E=210$ MPa) forms wrinkles when inflated with sufficient internal pressure, $P$. 
Crumples occur at lower pressure. 
Deflated bag width is $W=20$ cm. 
Supplementary Movies 1-3 show the transitions in panels (b-d), respectively. 
}
\label{fig:1}
\end{figure*}
\end{centering}

The realization that both wrinkles and crumples can form sequentially under gradual confinement was the outcome of recent work by King \textit{et al.}~\cite{King12,King13}. 
In their experiments, an initially-flat circular sheet is placed on a spherical water meniscus. 
As the curvature of the interface is gradually increased, a wrinkled state gives way to a different deformation mode with a finite number of stress-focusing patterns (Fig.~\ref{fig:1}a). 
This progression reveals two distinct symmetry-breaking transformations: First, wrinkles break the axial symmetry of the deformation field, and crumples subsequently break the axial symmetry of the stress field \cite{King12}. 
Yet, despite the practical importance and fundamental nature of the crumpling transition, a predictive understanding has remained out of reach.

Here we study the wrinkle-to-crumple transition in a set of model experiments on ultrathin polymer films and macroscopic inflated membranes. 
First, we show that this transition appears to be generic: 
We observe a strikingly similar morphological transition in hyperbolic and cylindrical geometries. 
Moreover, by isolating the wrinkle-to-crumple transition in a membrane inflated with gas, we show that the phenomenon is scale-free, and we identify robust morphological features of the crumpled phase that are shared across a diverse range of setups. 

We then characterize the crumpling threshold. 
We cross this transition by varying the curvature along the wrinkle crests in the interfacial films, and by modifying the tensile stresses along wrinkles in the inflated membranes. 
We generalize an empirical threshold from previous work \cite{King12} to give an approximate criteria for the transition, and we show that a full account of the crumpling threshold must also include the fractional in-plane compression. 

Wrinkles, folds, creases, ridges, blisters, and other buckled microstructures have been studied extensively in recent years~\cite{Vella09,Reis09,Kim11,Li12,Brau13,Rodriguez-Hernandez15,Wang16}. 
Crumple formation and evolution have not been documented to a similar extent and are still poorly understood. 
Our experimental measurements and phenomenological description provide a foothold for a theoretical understanding of this ubiquitous transition.

\section{Isolating the crumpling transition in diverse settings} 

We conduct experiments using interfacial films in two geometries that differ from the spherical setup where the transition was previously reported~\cite{King12}. 
We make polymer films of thickness $40 < t < 630$ nm and Young's modulus $E = 3.4$ GPa by spin-coating solutions of polystyrene ($M_\text{n} = 99$k, $M_\text{w} = 105.5$k, Polymer Source) in toluene ($99.9\%$, Fisher Scientific) onto glass substrates, following Ref.~\citealp{Huang07}. 
We cut the films into various shapes and float them onto deionized water with surface tension $\gamma = 72$ mN/m. 
We use a white-light interferometer (Filmetrics F3) to measure film thickness, which was uniform over each film to within $3\%$. 
Our experiments fall in a regime characterized by weak tension, $\gamma/Y < 10^{-3}$, and negligible bending stiffness, characterized by large ``bendability''~\cite{King12}, $\epsilon^{-1} = \gamma W^2/B > 10^{4}$, where $Y=Et$ and $B = Et^3/ [12 (1-\Lambda^2)]$ are the stretching and bending moduli, with $\Lambda$ the Poisson's ratio. 
Such films can withstand only vanishingly small levels of in-plane compression before they buckle out of plane. 

\begin{table*}[t]
\begin{tabular}{l l c c c c}
\hline
\hline
\textit{Setup}		& \textit{Control parameter}				& $\ell_\parallel$			& $R_\parallel$						& $\sigma_\parallel$					\\ 
\hline
Sheet-on-droplet	& Droplet radius ($R$)					& $W/2$ \text{(at transition)}			& $R/2$						& $\gamma$					\\
Indentation		& Indentation depth ($\delta$)				& $W$ 				& 2.62 $\ell_\text{curv}^2/\delta$ 	& $\gamma W/\ell_\text{curv}$ 		\\
Sheet-on-cylinder	& Radius ($R$), compression ($\Delta$) 		& $W$ 				& $R$					& $\gamma$ 					\\
Inflated membrane\ \ \ \ & Pressure ($P$)		 				& \ \ \ measured \ \ \ 		& \ \ \ \ \ measured \ \ \ \ \ 			& $P R_\parallel$ 					\\
\hline
\hline
\end{tabular}
\vspace{1cm}
\caption{
\textbf{Physical scales near the wrinkle-to-crumple transition.} 
Expressions for the buckled length ($\ell_\parallel$), curvature along the wrinkles ($R_\parallel$), and tensile stress along the wrinkles ($\sigma_\parallel$), which are pictured schematically in Fig.~\ref{fig:2}. 
Values for sheet-on-droplet are based on Ref.~\citealp{King12}. 
In this setup, $R_\parallel$ and $\sigma_\parallel$ vary spatially; we show their values at the location $r=W$ where the curvature is largest. (The full radial dependance may be obtained from Eqs.~24,25 in the SI to Ref.~\citealp{King12}.) 
Values for indentation are based on the height profile, $\zeta(r)$, in the relevant regime where wrinkles cover the sheet. 
In that case it was shown that $\zeta(r) = \delta \Ai(r/\ell_\text{curv})/\Ai(0)$, where $\Ai(x)$ is the Airy function and $\ell_\text{curv}=W^{1/3}\ell_\text{c}^{2/3}$ with $\ell_\text{c} = \sqrt{\gamma/\rho g}$ being the capillary length \cite{Vella15,Paulsen16}. 
The curvature $R_\parallel^{-1}(r) = \zeta''(r)$ is nearly maximal at $r=\ell_\text{curv}$, so we take $R_\parallel$ and $\sigma_\parallel$ there. 
For the inflated membranes, $\sigma_\parallel \approx P R_\parallel$ comes from force balance on a small cylindrical patch, $R_\parallel$ is measured using a set of paper stencils of circular arcs, and $\ell_\parallel$ is measured by laying a string along the buckled region \cite{bag-endnote}. 
}
\label{tab:1}
\end{table*}

Our first setup, shown in Fig.~\ref{fig:1}b and Supplementary Movie 1, is an indentation protocol that has been investigated previously \cite{Holmes10,Vella15,Paulsen16,Ripp18}. 
We indent a circular film of radius $11 < W < 44$ mm by a vertical distance $\delta$ using a spherical probe. 
At a threshold $\delta$, wrinkles form within a narrow annulus due to the azimuthal compression that would have been induced by the contraction of circles. 
The wrinkled region grows with increasing $\delta$ until it covers the sheet \cite{Vella18}. 
Beyond another threshold, some wrinkles increase in amplitude and develop into crumples, while the amplitude in the intervening regions decreases. 
The transition resembles the response in Fig.~\ref{fig:1}a, despite the markedly different geometry and loading. 
Here the small-scale wrinkles and crumples help the gross shape of the sheet (i.e., the overall geometry of the sheet that ignores undulations) to follow a horn-like profile with negative Gaussian curvature. 
We will return to analyze this and the following setups in Sections~\ref{sec:morph}-\ref{sec:threshold}. 

Figures~\ref{fig:1}a,b leave open the possibility that geometric incompatibility plays an important role in crumple formation. 
However, a crumpling transition can also be observed in a cylindrical geometry, where the Gaussian curvature of the gross shape is conserved while a principal curvature is made to vary. 
This is the character of the experiment shown in Fig.~\ref{fig:1}c and Supplementary Movie 2, where a rectangular polymer film of width $1.1 < W < \unit{3.2}{\milli\metre}$, and length $12 < L < 13$ mm is placed on a water meniscus that is pinned along two straight teflon walls that are $\unit{4.0}{\milli\metre}$ apart. 
First, we buckle the film by compressing it between two barriers using a micrometer stage. 
Then, by changing the height of a liquid reservoir, we vary the curvature of the interface continuously between $0$ (planar) and $0.50$ mm$^{-1}$ (half a cylinder). 
We confirmed that the meniscus shape shows no measurable deviations from a cylinder, both with and without a film on its surface, by measuring its profile with a sheet of laser light. 
We thus measure the curvature of the meniscus via its height in side-view images. 
The buckled sheet forms parallel wrinkles that transition into crumples beyond a threshold meniscus curvature (Fig.~\ref{fig:1}c). 
This is perhaps the minimal geometry for producing crumples, which we have done by pairing curvature and compression along perpendicular axes. 

To further probe the generality of this transition, we perform experiments where we quasistatically inflate sealed plastic membranes while measuring their internal pressure (Fig.~\ref{fig:1}d and Supplementary Movie 3). 
Square membranes of width $10 < W < 31$ cm and thickness $15 < t < 222$ $\mu$m are made by folding a rectangular sheet in half and sealing the three open sides. 
We use a variety of materials, including low-density polyethylene, perfluoroalkoxy alkane (PFA), polyolefin shrink film (SYTEC MVP), aluminized mylar, and natural rubber, in order to vary the Young's modulus over a wide range ($2.0 < E < 1500$ MPa), which we measure using a tensile tester (TestResources Model 100P). 
As the membrane is inflated, geometric constraints lead to buckling in four regions along the perimeter. 
We image the buckled zone on the side of the membrane without a seal, while measuring the internal pressure with a digital pressure gauge. 
At low gauge pressure, crumples are visible in this region; at higher pressure they transition into smooth wrinkles (Fig.~\ref{fig:1}d). 
Observing crumples in this setting without a liquid suggests that they are general features that arise out of a minimization of elastic energies in the sheet plus a substrate energy that helps impose the gross shape.

\begin{centering}
\begin{figure}[b]
\includegraphics[width=4.0cm]{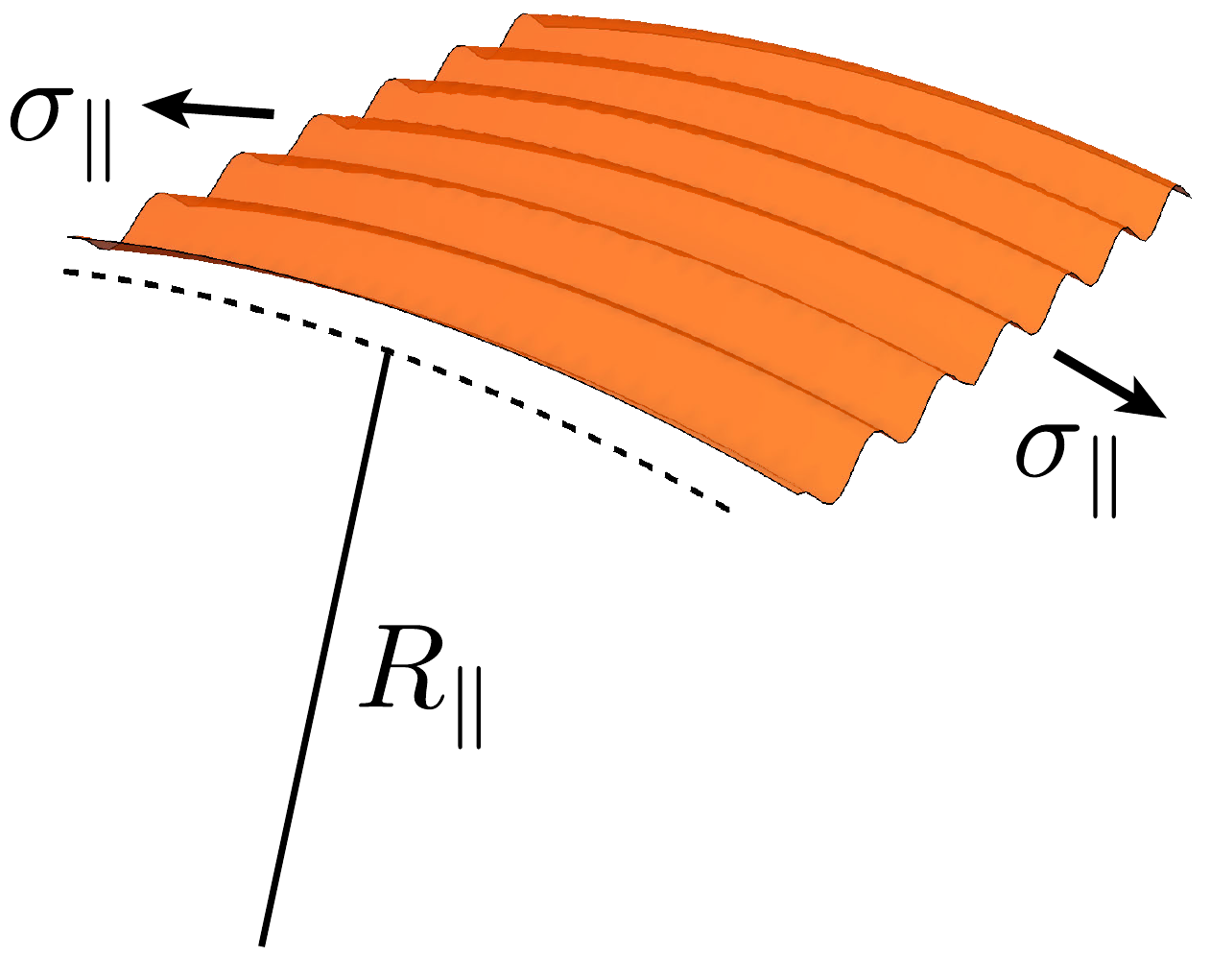} 
\caption{
Schematic of a wrinkled rectangular patch, showing $\sigma_\parallel$ and $R_\parallel$. Their values, and the total length of the buckled region, $\ell_\parallel$, are quantified in Table~\ref{tab:1} for each setup.
}
\label{fig:2}
\end{figure}
\end{centering}

\begin{centering}
\begin{figure*}[t]
\includegraphics[width=11.75cm]{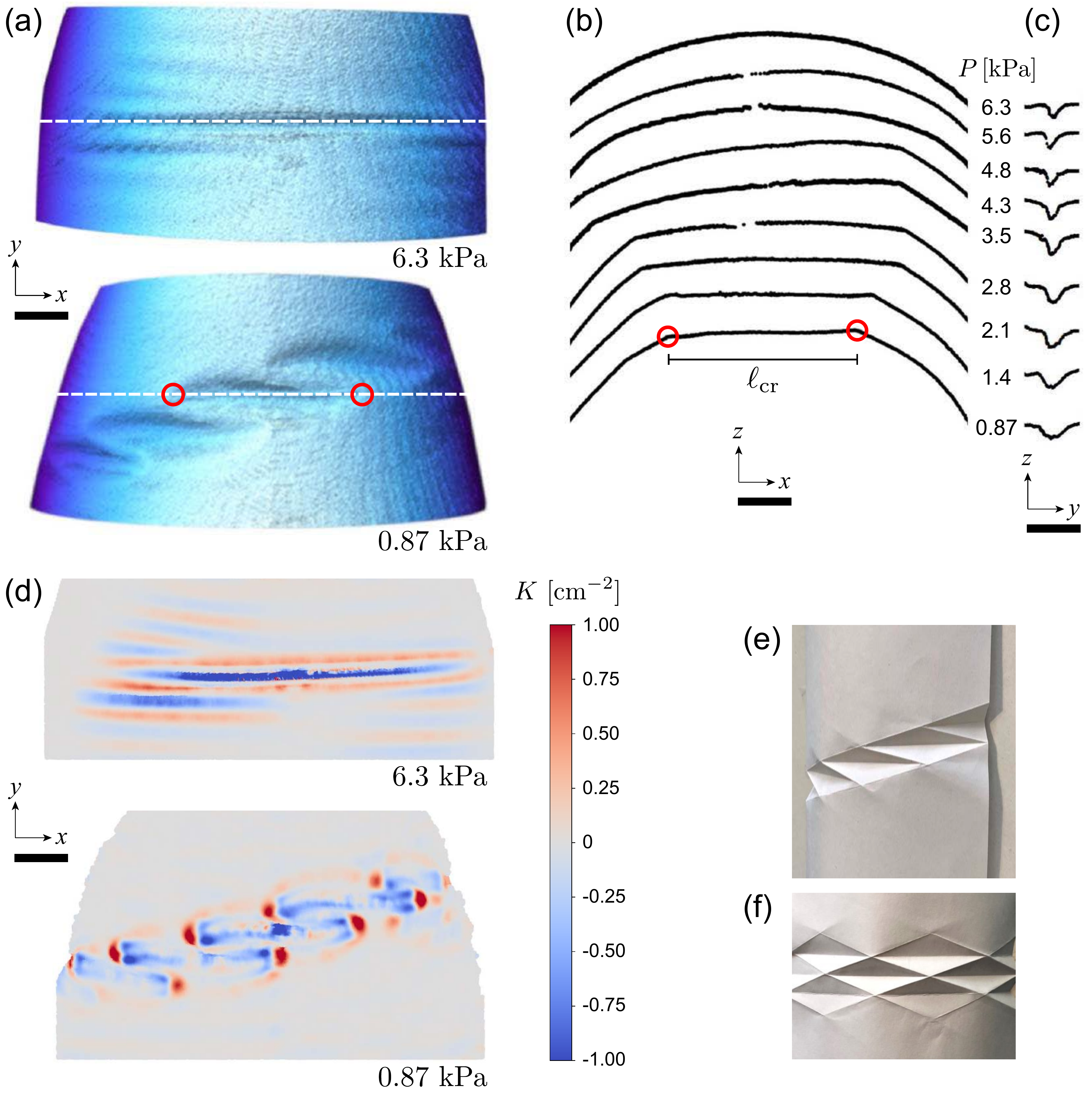} 
\caption{
\textbf{Crumple morphology.} 
\textbf{(a)} Renderings of three-dimensional scans of an inflated square polyethylene bag ($t=49$ $\mu$m, $E=297$ MPa, $W=20$ cm), showing the topography of wrinkles (top) and crumples (bottom). 
\textbf{(b)} Cross sections through a single buckled feature as a function of increasing pressure, taken along the dashed lines in panel (a). 
The two circles highlight the sharp tips at the two ends of the crumple. 
\textbf{(c)} Cross sections along a perpendicular direction, which only show a small change in width through the transition [same scale as (a,b)]. 
\textbf{(d)} Local Gaussian curvature, $K$, for the scans shown in panel (a). 
The wrinkled state at higher pressure (top) shows large positive and negative curvatures along the wrinkle crests and troughs. 
The crumpled state at lower pressure (bottom) localizes $K$ and hence material stresses. Scale bars in (a-d): 1 cm. 
\textbf{(e,f)} Origami patterns resembling crumpled states, suggesting that crumples may be an approximate isometry of the plane, with localized bending and stretching in place of the edges and vertices in these paper models. 
}
\label{fig:3}
\end{figure*}
\end{centering}

In order to draw quantitative comparisons between these four setups, we define a general set of variables for the region of the sheet that is wrinkled or crumpled. 
We denote the length of this region by $\ell_\parallel$, as drawn in Fig.~\ref{fig:1}c, which corresponds to the length of wrinkles when they are present. 
An individual ``crumple'' has a length $\ell_\text{cr} < \ell_\parallel$ (also indicated in Fig.~\ref{fig:1}c). 
We denote the radius of curvature and stress along the tensile direction by $R_\parallel$ and $\sigma_\parallel$, respectively, as pictured in Fig.~\ref{fig:2}. 
Table~\ref{tab:1} shows estimates for these variables in each setup. 
We use these variables to quantify the threshold for crumple formation in Section~\ref{sec:threshold}. 
But first, in Section~\ref{sec:morph}, we clarify the morphology of the crumpled state and how it evolves with increasing confinement.

\section{Crumple morphology}\label{sec:morph}

\subsection{Topography of a single crumple}\label{morph:1}

To define the difference between wrinkles and crumples more precisely, we use a laser scanner to map the three-dimensional topography of a portion of an inflated membrane at several internal pressures. 
Figure~\ref{fig:3}a shows renderings of the reconstructions at $P=6.3$ kPa where the sheet is wrinkled, and $P=0.87$ kPa where there are crumples, with $P$ being the gauge pressure. 
Figure~\ref{fig:3}b shows cross sections taken along the central trough, which we track through a series of pressures during a single inflation. 
At low pressure, the crumple trough is approximately flat and terminates at two localized regions of high curvature (red circles in Figs.~\ref{fig:3}a,b).   
As the internal pressure increases, these kinks gradually become smoother while the crumple gets longer until it spans the length of the buckled region; the crumple thereby converts into a smooth wrinkle. 

Figure~\ref{fig:3}c shows the evolution of the same buckled feature along a perpendicular cross section, taken halfway between the crumple tips. 
This undulation is present through the transition, and it serves to collect excess material due to lateral compression. 

\subsection{Gaussian curvature}

The observed wrinkle and crumple morphologies are linked to the distribution of stresses in the sheet via Gauss's \textit{Theorema Egregium}. 
To gain insight into the stress field, we measure the local Gaussian curvature of the scans by fitting small regions 
to a quadratic polynomial and extracting the principal curvatures from the coefficients of the fit, following Ref.~\citealp{Chopin16}. 
We calibrated the scanner and analysis code on a smooth metal cylinder, which gave the radius of the cylinder with an accuracy of $1\%$ and yielded spatial fluctuations in the Gaussian curvature on the order of $0.01$ cm$^{-2}$, which is one indicator of the measurement precision. 

Figure~\ref{fig:3}d shows the measured curvature maps for the scans in Fig.~\ref{fig:3}a. 
The wrinkled morphology shows finite Gaussian curvatures in stripes of alternating sign. 
This is due to the nearly constant curvature along the $x$-axis and the alternating positive and negative curvature of the wrinkled profile along the $y$-axis, which together imply finite strains and stresses that are spatially extended. 
In contrast, the curvature map of the crumpled morphology suggests that the material stresses are reduced throughout much of the sheet, at the expense of higher stresses around the boundary of the individual crumples, most notably at their tips. 
These data support a picture where crumples lower the total elastic energy by condensing stresses to small regions in the sheet. 

\subsection{Similarity to origami bellows}

These observations may lead one to consider isometries of a cylinder. 
The Yoshimura pattern \cite{Yoshimura55,Schenk14} and the Kresling pattern \cite{Kresling96} are two origami constructions that are built from a repeated diamond-shaped unit cell consisting of two flat triangular faces sharing an edge. 
Simultaneously actuating all the folds leads to global axial compression of the original cylinder. 

Taking one row of diamonds from the Yoshimura pattern yields the structure shown in Fig.~\ref{fig:3}e, which resembles the morphologies in Figs.~\ref{fig:1}c,d and \ref{fig:3}a. 
Symmetric patterns may be constructed as well (Fig.~\ref{fig:3}f). 
This qualitative similarity may prompt one to ask whether the crumped phase is a kind of ``self-organized origami''~\cite{Mahadevan05} that occurs in order for the sheet to be approximately isometric to an uncompressed cylinder (or equivalently, to the initial planar state). 
An answer might be reached by assessing the elastic energy cost of the finite-curvature ridges that replace the infinitely-sharp origami folds. 
It should also address the energetic cost for matching the crumpled region to the cylindrical profile flanking its sides.

\subsection{Crumple length}\label{morph:2}

Focusing now on a single crumple, we examine the cross sections in Fig.~\ref{fig:3}b once again. 
One can see the crumple length, $\ell_\text{cr}$, in these profiles at low pressure, by noting the distance between the pair of sharp kinks. 
The mechanism selecting this crumple length is not yet known, and our first task is to identify the parameters that affect it \cite{origami-endnote}. 
The cross sections in Fig.~\ref{fig:3}b show that $\ell_\text{cr}$ grows with increasing pressure, suggesting that $\ell_\text{cr}$ depends on the tensile stress, $\sigma_\parallel$. 
Optical images of the sheet-on-cylinder and indentation setups show that $\ell_\text{cr}$ also depends on the radius of curvature along the wrinkles, $R_\parallel$. 

In order to build an empirical scaling relation for the crumple length, 
we gather images in the indentation, sheet-on-cylinder, and inflated membrane setups. 
Because these systems span multiple scales, we are sensitive to any dependence on the length of the buckled region, $\ell_\parallel$. 
We record the longest crumple in each image, since smaller crumples may be associated with boundary effects. 
In the sheet-on-cylinder setup, we also performed experiments with $\sigma_\parallel = \gamma = 36$ mN/m by using a surfactant (sodium dodecyl sulfate). 

\begin{centering}
\begin{figure}[t]
\includegraphics[width=8.25cm]{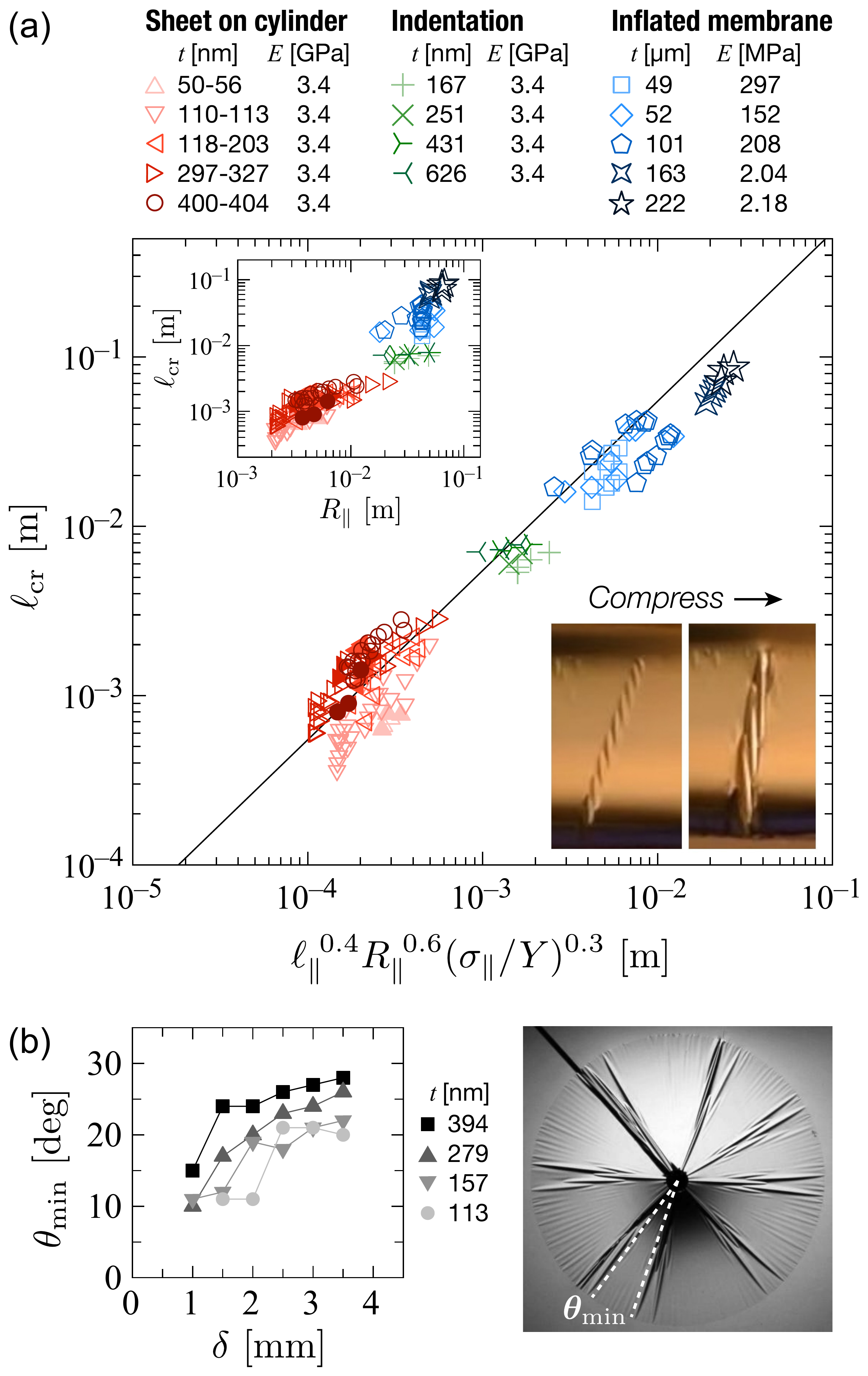}
\caption{
\textbf{Emergent scales in the crumpled phase.} 
\textbf{(a)}
Crumple length, $\ell_\text{cr}$, measured from optical images in the sheet-on-cylinder, indentation, and inflated membrane setups. 
Sheet thickness and modulus were varied over a wide range by using rubber ($E \sim 2$ MPa), polyethylene ($E \sim 200$ MPa), and polystyrene ($E = 3.4$ GPa) sheets. 
We also varied the surface tension in the sheet-on-droplet setup (open symbols: $72$ mN/m; closed symbols: $36$ mN/m). 
The data are reasonably described by a simple expression involving $\ell_\parallel$, $R_\parallel$, and $\sigma_\parallel/Y$ (solid line: Eq.~\ref{eq:ellcr}). 
The two images show $\ell_\text{cr}$ increasing upon compression; this may account for some of the scatter in the data. 
\textit{Inset:} $\ell_\text{cr}$ versus $R_\parallel$, which does not collapse the data. 
\textbf{(b)} 
Minimum angular separation between crumpled regions in indentation. 
The angular spacing grows with indentation depth, $\delta$, and is larger for thicker films. 
}
\label{fig:4}
\end{figure}
\end{centering}

Possible scaling relations for $\ell_\text{cr}$ are constrained by dimensional analysis and an observation from our experiments: 
At the crumpling transition, the crumple length is comparable to the wrinkle length, i.e., $\ell_\text{cr} \approx \ell_\parallel$. 
Based on the crumpling threshold presented in the following section (Eq.~\ref{eq:alpha}), we therefore expect that: $\ell_\text{cr} \sim \ell_\parallel^{(1-2\beta)} R_\parallel^{2\beta} (\sigma_\parallel/Y)^{\beta}$, for some $\beta$. 
Our measurements over two decades in $\ell_\text{cr}$ are reasonably described by: 
\begin{equation}\label{eq:ellcr}
\ell_\text{cr} \approx \text{5.6\ } \ell_\parallel^{\ 0.4} R_\parallel^{\ 0.6} \left( \frac{\sigma_\parallel}{Y} \right)^{0.3}, 
\end{equation}
as shown in Fig.~\ref{fig:4}a, which corresponds to $\beta=0.3$. 
Notably, the numerical prefactor in Eq.~\ref{eq:ellcr} is set by the above arguments, 
so that $\beta$ is the only parameter we fit to arrive at this result. 
Our data could also be consistent with $\beta$ ranging from $0.2$ and $0.4$. 
We were not able to produce a significantly better collapse by fitting the three exponents separately. 

The above scaling relation does not have any explicit dependance on the lateral compression applied to the sheet, which may explain some of the scatter in the data. 
Indeed, the images in Fig.~\ref{fig:4}a show a chain of crumples evolving under increasing compression, which causes $\ell_\text{cr}$ to grow. 
Thus, Eq.~\ref{eq:ellcr} is only approximate and should be modified to include a dependence on the compression. 
Nevertheless, the ability of Eq.~\ref{eq:ellcr} to capture our results from three very different systems supports our approach of describing the data using the general set of variables in Table~\ref{tab:1}.

\subsection{Lateral spacing between crumpled regions}\label{morph:3}

Looking now to the structure of the crumpled phase on a larger scale, we note that individual crumples organize into extended structures that span the length of the buckled region, as shown in each of the setups in Fig.~\ref{fig:1}. 
In some cases these structures have bilateral symmetry, like those in Figs.~\ref{fig:1}a,b, and the origami pattern in Fig.~\ref{fig:3}f. 
In another common motif, each crumple tip is situated near the midline of a neighboring crumple, as in Figs.~\ref{fig:1}c,d, and the origami pattern in Fig.~\ref{fig:3}e. 
We call these latter structures ``chains''. 
Both the symmetric and chain assemblies break the translational symmetry of the sheet-on-cylinder setup, whereas chains further break reflection symmetry due to their angled structure. 
Left- and right-angled chains can occupy the sheet at the same time, in equal or unequal numbers. 

These assemblies are ordered on a larger scale as well: they exhibit a relatively regular lateral spacing, with low-amplitude wrinkles or flat regions between them. 
This lateral spacing can be seen in Figs.~\ref{fig:1}b,c (denoted by $\ell_\perp$). 
It is also evident in the last panel of Fig.~S9 of Ref.~\citealp{King12} for the sheet-on-droplet setup. 
Because there is some stochastic variation in this lateral spacing, we focus on the minimum spacing within a sheet in the indentation setup, which gives the most reproducible results. 
In particular, we revisit the experiments of Paulsen \textit{et al.}~\cite{Paulsen16}, where a sheet of radius $W=11$ mm was indented from below. 
Figure~\ref{fig:4}b shows the minimum angular size $\theta$ of the region between crumples in a given sheet, which we measured for a variety of thicknesses and indentation amplitudes, $\delta$. 
This spacing is systematically larger for thicker films, and it grows with increasing $\delta$ as entire assemblies of crumples ``unfurl'', giving their excess azimuthal length to other areas of the sheet \cite{unfurl-endnote}. 

These data suggest that the spacing between crumpled regions is a robust feature of a uniform elastic sheet under suitable confinement, rather than an artifact of material defects in the film, which we expect would not lead to such systematic trends. 
Indeed, a characteristic angular separation is also evident in numerical simulations of an indented pressurized elastic shell (see Fig.~4d of Ref.~\citealp{Taffetani17}). 
Elucidating its underlying mechanism in each of the geometries remains an open challenge.

\section{Crumpling threshold}\label{sec:threshold}

\subsection{Confinement parameter}\label{sec:alpha}

Having characterized the basic phenomenology and morphology of the crumpled state, we now give a quantitative account of the crumpling threshold. 
As we will show, the crumpling transition in these four experimental setups may be gathered in a single empirical phase diagram. 
For the case of a thin circular sheet on a droplet, King \textit{et al.}~\cite{King12} identified a basic dimensionless group governing the morphological transitions seen in experiment, given by $\alpha \equiv Y W^2/(2\gamma R^2)$, where $R$ is the radius of curvature of the droplet \cite{King12}. 
This expression may be seen as a ratio of geometric strain ($\sim$$W^2/R^2$), to mechanical strain ($\sim$$\gamma/Y$). 
For the experiment in Fig.~\ref{fig:1}a, King \textit{et al.}~predicted the appearance of wrinkles when $\alpha = \alpha_\text{wr} \approx 5.16$, consistent with their experiments, and they observed a theoretically-unanticipated crumpling transition when $\alpha = \alpha_\text{cr} \approx 155$. 
To generalize this empirical threshold so that it may be compared with other setups, we replace $\gamma$, $R$, and $W$ with the general set of variables introduced in Table~\ref{tab:1}. 
Thus,
\begin{equation}\label{eq:alpha}
\alpha \equiv \frac{Y \ell_\parallel^2}{2\sigma_\parallel R_\parallel^2} \ . 
\end{equation}

\begin{centering}
\begin{figure}[t]
\includegraphics[width=8.75cm]{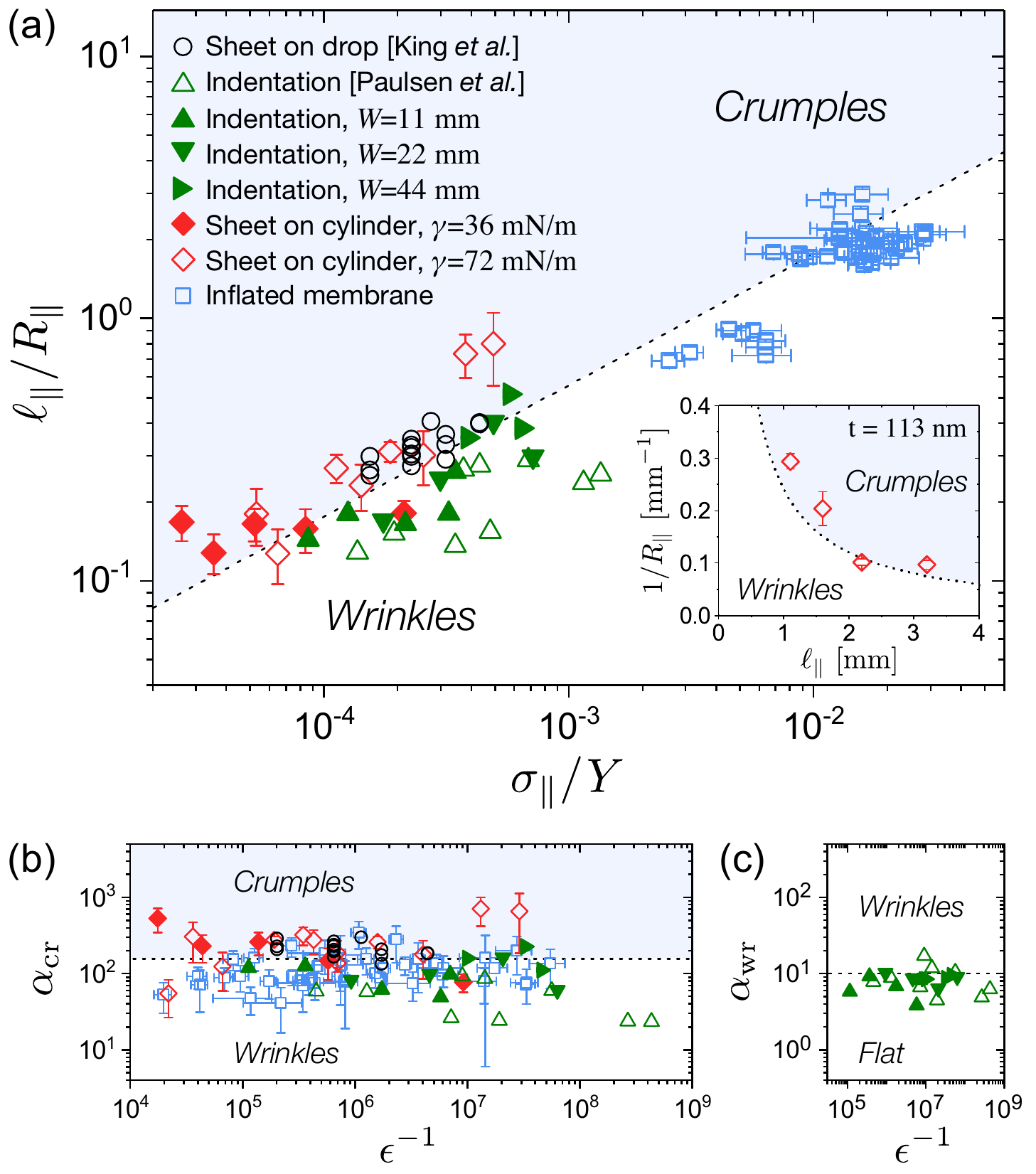}
\caption{
\textbf{Phase diagram for wrinkles and crumples.} 
\textbf{(a)} 
Crumpling threshold measured in the four experimental setups in Fig.~\ref{fig:1}, including sheet-on-droplet data from King \textit{et al.}~\cite{King12} and indentation data from Paulsen \textit{et al.}~\cite{Paulsen16}. 
Values of $\ell_\parallel$, $R_\parallel$, and $\sigma_\parallel$ are either measured or deduced from other measured quantities (see Table~\ref{tab:1}). 
Sheet thickness ranges from $40 < t < 630$ nm for the polymer films and $15 < t < 222$ $\mu$m for the inflated membranes. 
Indentation data have $\gamma = 72$ mN/m and $W$ shown in legend. 
Sheet-on-cylinder data have $W = 3.2$ mm and $\gamma$ shown in legend. 
Parameters for inflated membranes are detailed in Fig.~\ref{fig:6}a,b. 
The data are reasonably well-described by Eq.~\ref{eq:alpha} with $\alpha=\alpha_\text{cr}=155$ (dashed line). 
\textit{Inset:} Threshold curvature $1/R_\parallel$ for crumpling, versus wrinkle length $\ell_\parallel$, in the sheet-on-cylinder setup at fixed $t$ and $\gamma$. 
Dashed line: Eq.~\ref{eq:alpha} with $\alpha = 155$. 
\textbf{(b)} Crumpling threshold versus bendability, $\epsilon^{-1}$. 
The threshold is approximately constant over a wide range of bendability. 
\textbf{(c)} Wrinkling threshold in the indentation setup, measured in the same indentation experiments as panels (a,b). 
The scatter in the filled and open triangles are similar in magnitude to their scatter in panel (b), respectively. 
}
\label{fig:5}
\end{figure}
\end{centering}

\begin{centering}
\begin{figure*}[t]
\includegraphics[width=15.75cm]{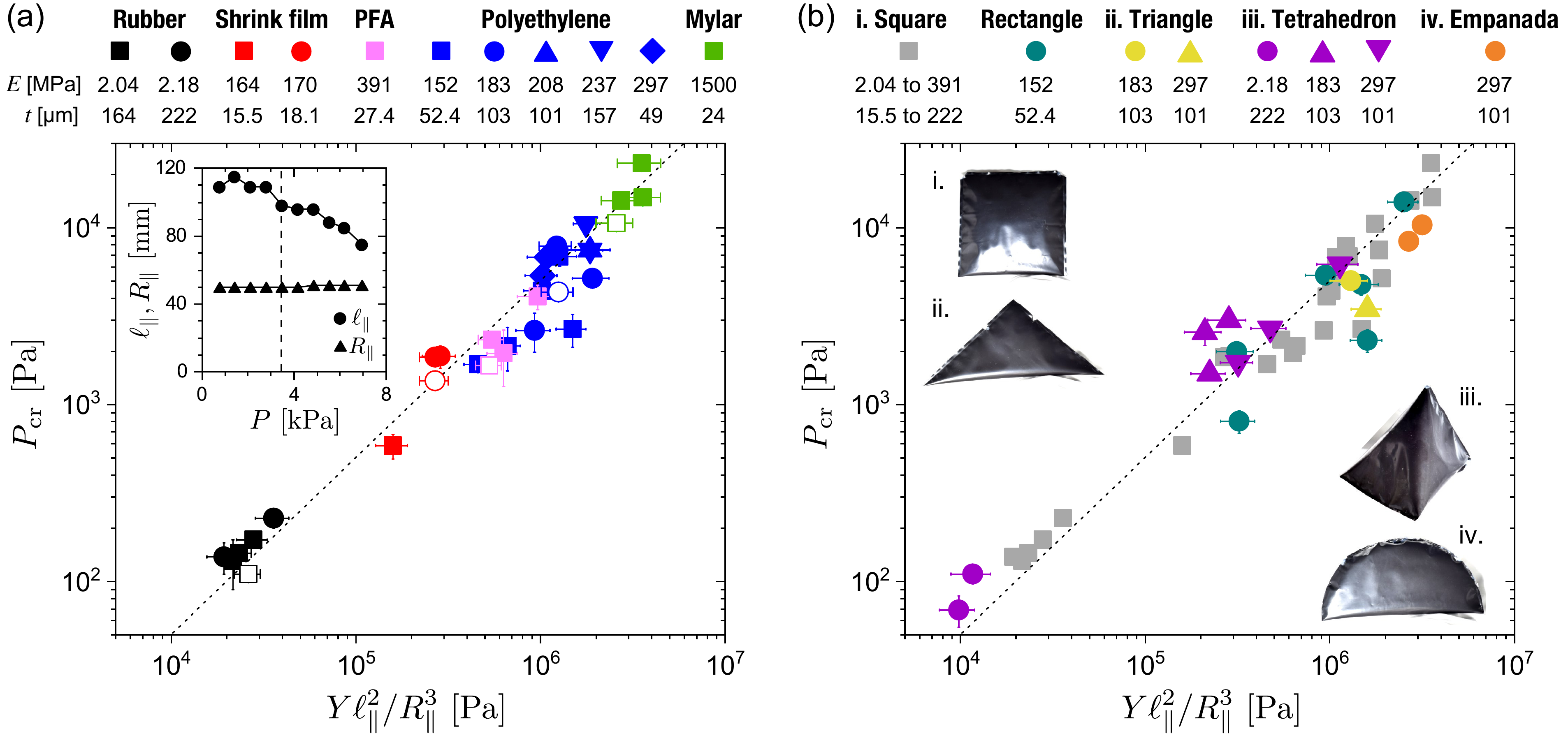}
\caption{
\textbf{Pressure threshold for transitioning from sharp crumples to smooth wrinkles in inflated membranes.} 
\textbf{(a)} 
Experiments using square bags of various thickness ($15 < t < 222$ $\mu$m), width ($10 < W < 31$ cm), and Young's modulus ($2.0 < E < 1500$ MPa). 
Open symbols show the transition for decreasing pressure; the rest of the data were obtained by increasing the pressure. 
The data are well-described by Eq.~\ref{eq:pressure} with $\alpha_\text{cr} = 100$ (dashed line). 
Inset: $\ell_\parallel$ and $R_\parallel$ versus pressure for a square polyethylene bag ($W=20$ cm, $t=49$ $\mu$m). 
Both quantities are relatively constant near the threshold pressure (dashed line: $P_\text{cr}$). 
\textbf{(b)} 
Experiments with different bag shapes, which are shown in the insets prior to inflation. 
We used a variety of polyethylene ($E \sim 200$ MPa) and rubber ($E \sim 2$ MPa) sheets to vary the Young's modulus and thickness, as denoted in the legend. 
Here, $9.4 < W < 31$ cm. 
The crumpling threshold is consistent with the result for square bags, as shown by the gray squares and dashed line that are repeated from panel (a). 
}
\label{fig:6}
\end{figure*}
\end{centering}

Figure~\ref{fig:5}a shows the crumpling thresholds measured in the four setups, plotted as a function of $\ell_\parallel/R_\parallel$ and $\sigma_\parallel/Y$, which characterize the magnitude of the imposed curvature and tensile strains, respectively. 
(The rightmost points indicate that the bags may undergo macroscopic strains of order $3\%$ at the transition; the polymer films experience significantly smaller strains.) 
For the experiments using floating films, the threshold is traversed vertically by varying the imposed curvature, $1/R_\parallel$. 
For the inflated membranes, the threshold is crossed horizontally, as changing the internal pressure causes $\sigma_\parallel$ to vary while $\ell_\parallel/R_\parallel$ remains approximately constant (see the inset to Fig.~\ref{fig:6}a). 
The dashed line shows $\alpha_\text{cr} = 155$. 
Although there is significant scatter (corresponding to $\alpha_\text{cr}$ ranging from $23$ to $700$), this simple scaling organizes the data into a single phase diagram, covering a wide range of experimental setups with different sheet geometries and confinement protocols. 

To probe this picture further, we examine the dependence of the crumpling threshold on the buckled length, $\ell_\parallel$. 
Equation~\ref{eq:alpha} implies that longer wrinkles should transition into crumples at smaller curvature, $1/R_\parallel$. 
This dependence may be tested directly in the sheet-on-cylinder setup, where the buckled region spans the entire sheet width, i.e., $\ell_\parallel = W$. 
We perform experiments with a single sheet thickness, $t=113$ nm, and vary the sheet width, $1.1 < W < \unit{3.2}{\milli\metre}$. 
The inset to Fig.~\ref{fig:5}a shows the observed curvature where crumples appear. 
The data are consistent with $1/R_\parallel \propto 1/\ell_\parallel$ at the crumpling transition, following Eq.~\ref{eq:alpha} with $\alpha=155$. 

King \textit{et al.}~\cite{King12} proposed that the crumpling threshold is independent of bendability, $\epsilon^{-1} = \gamma W^2/B$, in the asymptotic limit $\epsilon^{-1} \gg 1$, although their measurements were confined to two decades: $10^5 < \epsilon^{-1} < 10^7$. 
Our measurements greatly expand this range. 
Figure \ref{fig:5}b shows that $\alpha_\text{cr}$ is approximately constant for $10^4 < \epsilon^{-1} < 10^9$ in the floating polymer films and the inflated membranes (where we define $\epsilon^{-1} = P R_\parallel W^2 / B$, with $P$ the pressure at the crumpling transition). 


Although we have gleaned significant overall trends from these experiments, there is a fairly broad distribution of observed crumpling thresholds, $\alpha_\text{cr}$. 
We look at the scatter in another transition at smaller confinement as a basis for comparison. 
The wrinkling transition in indentation has been studied previously using theory and experiments \cite{Vella15,Vella18,Ripp18}; translating these results to our variables \cite{alpha-endnote} gives a predicted threshold of $\alpha_\text{wr} = 10.02$. 
Figure~\ref{fig:5}c shows experimental measurements that are clustered around this value, although many sheets undergo the transition at smaller $\alpha$. 
This behavior might be caused by imperfections in the sheet or at its edge, which could bring about wrinkles at smaller $\alpha$, and could likewise contribute to the scatter in the crumpling threshold. 

Note that there is an overall shift between the open and closed triangles in Fig.~\ref{fig:5}b that is absent in Fig.~\ref{fig:5}c. 
This is because the appearance of wrinkles is a sharp criteria that is easily identified in the images, whereas the crumpling transition is much broader. 
For indentation, Paulsen \textit{et al.}~\cite{Paulsen16} (open triangles) reported the beginning of the transition by identifying the growth of the wrinkle amplitude in discrete regions that eventually become crumples. 
In contrast, the filled triangles from the present study indicate where the sharp tips at the end of the crumples are clearly visible---an event that occurs at larger curvature.

\subsection{Pressure threshold for inflated membranes}\label{sec:bags}

The threshold confinement, $\alpha_\text{cr}$, for inflated membranes may be recast as a threshold pressure. 
Plugging $\sigma_\parallel \approx PR_\parallel$ into Eq.~\ref{eq:alpha}, we obtain: 
\begin{equation}\label{eq:pressure}
P_\text{cr} \approx \frac{1}{2 \alpha_\text{cr}}\frac{Y \ell_\parallel^2}{R_\parallel^3} \ . 
\end{equation}
The threshold $\alpha_\text{cr}$ may in principle depend on the shape of the bag, but a good estimate should be given by the value measured in the sheet-on-droplet experiments. 
Figure~\ref{fig:6}a shows the measured threshold pressure in experiments where we gradually increase the internal pressure. 
We varied the stretching modulus, $Y$, over $2$ orders of magnitude by constructing bags from different polymer or rubber sheets. 
The data are captured by Eq.~\ref{eq:pressure}, and we obtain $\alpha_\text{cr} = 100 \pm 30$ by fitting for the numerical prefactor. 

We also measured the transition upon decreasing the internal pressure in one bag of each material. 
The open symbols in Fig.~\ref{fig:6}a show these measurements, which are systematically lower than the points for increasing pressure. 
By cycling the pressure within each bag, we measured a ratio of transition pressures of $0.55$ for the polyethylene bag, and a ratio between $0.71$ and $0.76$ for the other four materials. 
We did not observe such a strong hysteresis in the polymer films (Fig.~\ref{fig:7}). 
One possible interpretation is that the larger strains imposed on the bags (estimated by the range of $\sigma_\parallel/Y$ shown in Fig.~\ref{fig:5}) could lead to larger plastic deformations in both the wrinkled and crumpled phases, thereby biasing the deformation pattern towards what is already there. 

We now move to test the generality of Eq.~\ref{eq:pressure} by constructing bags of different shapes, as pictured in Fig.~\ref{fig:6}b. 
Predicting the inflated shape of the bag or the locations of the buckled regions are both non-trivial tasks~\cite{Paulsen94, Deakin09, Barsotti14, Vetter14, Siefert19}. 
Nevertheless, we may simply measure $\ell_\parallel$ and $R_\parallel$ at the crumping transition. 
Plugging these measurements into Eq.~\ref{eq:pressure} gives a good estimate of the threshold pressure, $P_\text{cr}$ for each bag shape, size, material, and thickness, as shown in Fig.~\ref{fig:6}b. 
Moreover, our experiments show that $\ell_\parallel$ and $R_\parallel$ do not vary significantly as a function of pressure, as shown in the inset to Fig.~\ref{fig:6}a for a square polyethylene bag. 
Thus, one may obtain a basic estimate of the minimum pressure $P_\text{cr}$ required to replace sharp crumples with smooth wrinkles by measuring $\ell_\parallel$ and $R_\parallel$ at lower pressures. 

These results suggest that the bag geometry affects the crumpling threshold through a straightforward mechanism, i.e., by creating a compressive zone and selecting $\ell_\parallel$ and $R_\parallel$ there. 
However, as we will describe in Sec.~\ref{sec:compression}, experiments in the sheet-on-cylinder setup show that the crumpling threshold depends also on the compression, $\Delta$, which for the inflated membranes depends on the bag geometry and the internal pressure. 
This additional consideration implies a more nuanced coupling of the bag geometry to the crumpling transition, although it could still be consistent with Eq.~\ref{eq:pressure} with a threshold $\alpha_\text{cr}$ that depends on the compression, i.e., $\alpha_\text{cr} = \alpha_\text{cr}(\Delta)$. 
As we will now show, this general idea is supported by more detailed experiments in the sheet-on-cylinder setup.

\subsection{Disentangling curvature and compression}\label{sec:compression} 

\begin{centering}
\begin{figure}[t]
\includegraphics[width=8.5cm]{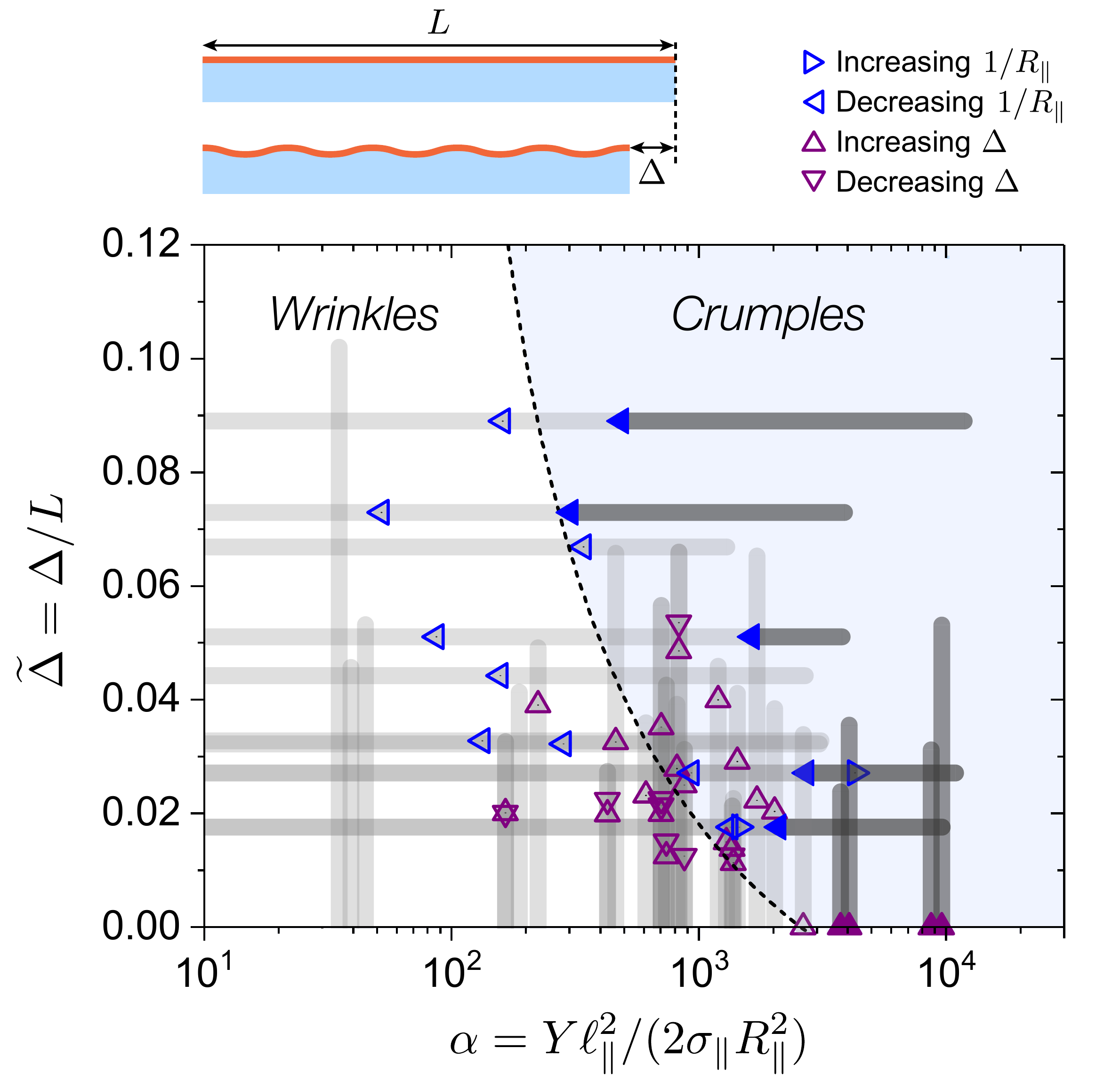}
\caption{
\textbf{Role of compression in crumpling threshold.} 
Symbols show the observed crumpling transition in experiments on the sheet-on-cylinder setup, where either the curvature, $1/R_\parallel$, or the compression, $\Delta$, were varied while the other was held fixed. 
The sheet width ($1.6 < W < 3.2$ mm) and thickness ($50 < t < 300$ nm) were also varied, and $\gamma=72$ mN/m. 
Gray bands show the range probed in experiments. 
Over an intermediate range of $\alpha$, wrinkles or crumples may both exist depending on the compression. 
Filled symbols denote a transition to a morphology with crumples but no wrinkles, marked by the darker gray bands. 
(Open symbols with fixed $\tilde{\Delta} > 0.03$ are also shown in Fig.~\ref{fig:5}.) 
}
\label{fig:7}
\end{figure}
\end{centering}

In the sheet-on-drop, indentation, and inflated membrane setups, compression is achieved by causing the sheet to approximate a surface with non-zero Gaussian curvature. 
In this way, the compression and curvature fields are linked---they are two complementary aspects of confining a planar sheet in a geometrically-incompatible setting. 
To separate the distinct roles of curvature and compression, we conduct further experiments in the sheet-on-cylinder setup. 
We situate a rectangular sheet of width $1.6 < W < 3.2$ mm and length $L = 12 \pm 0.5$ mm on a cylindrically-curved liquid meniscus of radius $R_\parallel$, and then quasistatically compress the film a distance $\Delta$ on its long axis by moving one of the barriers via a micrometer stage.  
Figure~\ref{fig:7} shows the observed crumpling transition as a function of the confinement parameter, $\alpha$, and the fractional compression, $\widetilde{\Delta} = \Delta/L$. 
For small $\alpha$, only wrinkles are observed for the entire range of $\widetilde{\Delta}$ probed. 
For intermediate $\alpha$, wrinkles are observed at small $\widetilde{\Delta}$ and crumples appear at large $\widetilde{\Delta}$. 
We thus observe a transition from wrinkles to crumples as a function of $\widetilde{\Delta}$. 
For large $\alpha$, we do not detect a wrinkled phase; to within our experimental resolution, crumples arise as soon as the sheet is compressed.

In another set of experiments, we first compress the film and then vary the curvature. 
We observe a transition between wrinkles and crumples at a threshold $\alpha$ that depends on the particular value of $\widetilde{\Delta}$ (right and left facing triangles in Fig.~\ref{fig:7}).  
The threshold is consistent with our experiments at fixed $\alpha$ and varying $\widetilde{\Delta}$ (up and down facing triangles), although there is significant run-to-run variation. 
We do not detect any systematic hysteresis across these experiments (in contrast to what was observed for the inflated membranes). 
In these experiments where the value of $\widetilde{\Delta}$ is held fixed, we witness two transitions as we increase the curvature, passing through three distinct buckling responses: (i) wrinkles only, (ii) wrinkles and crumples, and (iii) crumples only. 
[In keeping with previous terminology~\cite{King12}, we refer to both (ii) and (iii) as a ``crumpled'' phase.] 
In response (iii), compression is relieved only by crumples, separated by unbuckled regions. 
We mark the range where we observe this behavior with dark bands in Fig.~\ref{fig:7}. 
This response also occurs in experiments with fixed curvature when $\alpha$ is sufficiently large ($\alpha \gtrsim 3000$). 

These observations prompt us to ask how wrinkles and crumples may appear next to each other on the same sheet. 
One interpretation is that such ``coexistence'' can arise if the compression field is spatially varying. 
Thus, wrinkles or crumples occur where $\widetilde{\Delta}$ is locally small or large, respectively, for suitable intermediate $\alpha$. 
We emphasize that the vertical axis of Fig.~\ref{fig:7} is the \textit{global} compression. 
One could imagine an analogous phase diagram with \textit{local} compression as a vertical axis, where coexistence would correspond to different regions of the sheet occupying different positions along the vertical axis. 
This may also account for some of the scatter in the data in Figs.~\ref{fig:5},\ref{fig:7}, since the sheet can pass material between adjacent buckled regions. 
Namely, a wrinkled state could give way to a crumpled state by a small ``trade'' of excess length that pushes a region of the sheet to higher $\widetilde{\Delta}$ that exceeds the crumpling threshold. 

At sufficiently large $\widetilde{\Delta}$, the sheets are found to fold---they spontaneously gather material into a localized region where the film contacts itself. 
We limited our measurements in Fig.~\ref{fig:7} to sufficiently small $\widetilde{\Delta}$ where such folds do not appear. 
For the case of a flat bath ($\alpha=0$), there is a well-studied wrinkle-to-fold transition that results from a competition of bending and gravitational energies~\cite{Pocivavsek08,Brau13,Oshri15}, but this folding threshold depends on $\Delta$ rather than $\widetilde{\Delta}$, so we do not denote it in Fig.~\ref{fig:7}.

\section{Discussion}

We have shown that wrinkles are unstable to another buckled morphology at large curvatures, namely, sharp localized crumples. 
Although this transition was observed in previous experiments on circular polymer films in a spherical geometry~\cite{King12,King13}, we have shown that this symmetry-breaking event appears to be a generic phenomenon by isolating and characterizing the transition in a wide range of experimental setups across multiple lengthscales. 
These varied experimental geometries show that crumple formation is not unique to a particular overall Gaussian curvature; rather, crumples are sufficiently robust to form in spherical, hyperbolic, and cylindrical settings. 

By showing that a quantitatively similar transition occurs for both interfacial films and inflated membranes, our work suggests that a competition of elastic energies along with a suitable substrate energy is enough to give this rich behavior. 
This view is plausible given the basic observation that wrinkle crests must traverse a longer arclength than wrinkle troughs in curved topographies, implying costly stretching~\cite{Paulsen16}. 
Crumples may offer a lower-energy solution by condensing stresses to smaller regions on the sheet. 
Indeed, our topographic measurements suggest that at high confinement, a portion of a crumple trough may be nearly developable (i.e., isometric to the original planar sheet). 
In contrast, the region near a crumple tip has significant localized stretching. 
Whether this region is well-approximated by a d-cone \cite{Chaieb98,Cerda98,Boudaoud00,Walsh11,Gottesman15}, as suggested by King \textit{et al.}~\cite{King12,King13}, remains to be elucidated. 
Accounting for the elastic cost of these structures, in a manner that is consistent with various geometric constraints, could give insight into this transition. 

Another foothold for theoretical study comes from our results in the sheet-on-cylinder setup at large $\alpha$, which suggest a transition directly from a cylinder to a crumpled state, without an intervening wrinkled state. 
This opens up the possibility that at sufficiently large curvature, crumples might be modeled through an expansion around a smooth cylinder in analogy to what has been achieved for wrinkling in the far-from-threshold approach \cite{Davidovitch11,Hohlfeld15,Vella15,Davidovitch19}. 

In contrast to the shell buckling of an axially-loaded hollow cylinder~\cite{Donnell34,Karman41,Virot17}, the distinctive crumple morphology studied here relies on hoop tension. 
Cylinders subjected to static circumferential tension are common in industrial settings, for instance in liquid storage tanks. 
We note a striking visual similarity between crumples in the sheet-on-cylinder setup (Fig.~\ref{fig:1}c) and the localized buckling patterns that were observed on stainless-steel wine tanks due to an earthquake (see Fig.~1.7b in Ref.~\citealp{Bushnell81}). 
Crumples appear to be highly adaptable to a variety of geometries, as shown by the range of experimental setups explored here, and also the variety of membrane shapes. 
Qualitatively similar features have also been observed in pressurized spherical shells~\cite{Taffetani17} and rapidly-crushed conical shells~\cite{Gottesman18}, which raise questions about how intrinsic curvature and dynamics could influence stress focusing of the kind studied here. 

Many of the results presented in this article are suggestive of a local view of crumpling---that is, a description of crumpling in terms of the local values of a small set of variables. 
To summarize, we have shown how the crumpling threshold depends on the curvature and system size via $\ell_\parallel/R_\parallel$, the in-plane tensile strain via $\sigma_\parallel/Y$, and the fractional compression, $\Delta/L$, in phase diagrams spanned by combinations of these variables (Figs.~\ref{fig:5},\ref{fig:7}). 
We have also presented a picture wherein the local value of the compression is key to understanding the coexistence of wrinkles and crumples. 

Despite this promising local view that may capture important aspects of the transition, nonlocal effects may matter as well. 
In the axisymmetric setups studied here, the curvature and compression vary with the radial coordinate, but we observe chains of crumples appearing along radial lines all at once, rather than starting at a location of high curvature and growing in extent with increasing confinement. 
There are also hints of organization at intermediate scales in the crumpled phase: Chains of crumples seem to have a well-defined spacing between them ($\ell_\perp$ in Fig.~\ref{fig:1}c, and $\theta_\text{min}$ in Fig~\ref{fig:4}b). 
Although there are fluctuations in this spacing, it appears to have a reproducible minimum value, which is smaller for thinner sheets. 
These observations are suggestive of a domain structure~\cite{Aharoni17,Tovkach20} where chains of crumples are separated by regions of smaller-amplitude deformations as a repeated motif. 
This stands in contrast to the space-filling buckling patterns that are observed when a thin cylinder is axially compressed around a mandrel of slightly smaller diameter \cite{Horton65,Seffen14,Tobasco17}. 
Understanding the origin of this emergent mesoscopic lengthscale, and why it arises for a fluid substrate but not for a solid mandrel, could allow one to control these patterns with a suitably engineered substrate or sheet. 

On the micro-scale, wrinkles have been used for metrologies of films~\cite{Chung11}, for making smart surfaces with tunable wetting and adhesion~\cite{Chan08,Zang13}, and to conduct surface microfluidics~\cite{Khare09}. 
Our results expand the vocabulary of film deformations for advanced materials, and illustrate how buckled microstructures may change their nature in curved topographies. 
Discovering the mechanism of this symmetry-breaking instability remains an open challenge that should be the subject of future work.

\begin{acknowledgments}
We thank Hillel Aharoni, Julien Chopin, Benny Davidovitch, Vincent D\'emery, Narayanan Menon, Ian Tobasco, Zhanlong Qiu, Robert Schroll, and Dominic Vella for enlightening discussions. 
We thank Vincent D\'emery, James Hanna, Hunter King, Ian Tobasco, and Dominic Vella for useful comments on the manuscript. 
We thank the Syracuse Biomaterials Institute for use of a tensile tester.
This work was supported by NSF Grants No.~DMR-CAREER-1654102, No.~REU DMR-1757749, and No.~IGERT-1068780. 
\end{acknowledgments}


\begin{thebibliography}{69}
\bibitem [{bag()}]{bag-endnote}%
  \BibitemOpen
  \href@noop {} {}\bibinfo {howpublished} {For some square bags, $R_\parallel$
  was obtained by multiplying the height of the bag in the side-view images by
  a conversion factor that we determined for each material, and
  $\ell_\parallel$ was determined from the side-view images via its projection
  onto a circle of this radius.}\BibitemShut {Stop}%
\bibitem [{ori()}]{origami-endnote}%
  \BibitemOpen
  \href@noop {} {}\bibinfo {howpublished} {Note that although the origami
  bellows in Fig.~\ref{fig:3}e may be used to suggest geometric relationships
  between various lengthscales in the pattern, such a line of reasoning does
  not assign a unique crumple length to a given overall curvature,
  $R_\parallel$.}\BibitemShut {Stop}%
\bibitem [{unf()}]{unfurl-endnote}%
  \BibitemOpen
  \href@noop {} {}\bibinfo {howpublished} {The ability of these crumples to
  quickly disperse suggests that there is little or no self-contact in a
  crumpled region, as the polymer film would adhere to itself in that
  case.}\BibitemShut {Stop}%
\bibitem [{alp()}]{alpha-endnote}%
  \BibitemOpen
  \href@noop {} {}\bibinfo {howpublished} {Using the expressions in
  Table~\ref{tab:1}, one finds that $\alpha = \tilde{\delta}^2/(2 \cdot
  2.62^2)$, where $\tilde{\delta}$ is a dimensionless indentation depth defined
  in Ref.~\citealp{Vella15}.}\BibitemShut {Stop}%
\end{thebibliography}
\end{document}